\documentclass[%
 aip,
 amsmath,amssymb,
reprint,%
]{revtex4-1}
\usepackage{multirow}
\usepackage{mathrsfs}  
\usepackage[section]{placeins}
\usepackage{graphicx}% Include figure files
\usepackage{dcolumn}% Align table columns on decimal point
\usepackage{bm}% bold math
\usepackage[utf8]{inputenc}
\usepackage[T1]{fontenc}
\usepackage{mathptmx}
\usepackage{etoolbox}

\usepackage[dvipsnames]{xcolor}

\makeatletter
\def\@email#1#2{%
 \endgroup
 \patchcmd{\titleblock@produce}
  {\frontmatter@RRAPformat}
  {\frontmatter@RRAPformat{\produce@RRAP{*#1\href{mailto:#2}{#2}}}\frontmatter@RRAPformat}
  {}{}
}%

\makeatother
\begin{document}

%\preprint{AIP/123-QED}

\title{Insights into the electrorheological and electrohydrodynamic regimes in electrically driven emulsions}

\author{Majid Bahraminasr}

\affiliation{Department of Physics and Physical Oceanography, Memorial University, St. John’s, NL A1B3X7, Canada 
}
\altaffiliation[Also at ]{Department of Physics, University of Guelph, Guelph, ON N1G 2W1, Canada
}%Lines break automatically or can be forced with \\
\author{Anand Yethiraj}%
\homepage{https://softmaterials.ca/}
 \email{ayethira@uoguelph.ca}
\affiliation{ Department of Physics, University of Guelph, Guelph, ON N1G 2W1, Canada
}%

\date{\today}% It is always \today, today,

\begin{abstract}
Recently, we reported the electrorheoimaging (ERI) technique (Bahraminasr et al, 2026), and found that frequency-dependent electric field of an oil-in-oil emulsion yields two distinct regimes: a high-frequency dipolar, electrorheological (ER) regime and a low-frequency electrohydrodynamic (EHD) regime. In this work, we identify a phenomenological model to fit the results in the ER regime to a classic yield-stress fluid, and find collapse onto a master curve upon rescaling, consistent with a yield stress that grows approximately as $E^2$.
Macroscopic small-amplitude oscillatory shear (SAOS) rheology is compared with passive microrheology employing differential dynamic microscopy (DDM), with the close agreement implying scale independence of the ER behaviour, and indicating that, unlike steady shear, SAOS measurements do not restructure these samples and probe underlying material properties. 
Finally, under the presence of both steady shear and electric fields in the EHD regime, the emulsion forms banded structures composed of alternating droplet-rich and droplet-depleted regions. We explore recurrence and divergence in the location of these bands: they emerge within seconds of field application and decay rapidly after the field is switched off. Using the Jensen--Shannon divergence between radial intensity profiles, we show that the driven structure loses memory on timescales of order $1~s$ commensurate with the timescale of the EHD convection roll. For much longer field-off intervals successive banding events become statistically independent. 
\end{abstract}

\maketitle

\section{\label{sec:intro}Introduction}

Electrorheological (ER) fluids -- suspensions of electrically polarizable particles or droplets dispersed in an suspending liquid \cite{winslow1949induced, kesy2022characteristics} -- undergo rapid and reversible microstructural rearrangements, that lead to tunable mechanical properties that can be controlled by an external electric field~\cite{dong2019recent,liang2023efficient}. 
Electrically driven emulsions provide a class of ER-like systems in which electrohydrodynamic (EHD) stresses act on deformable interfaces, 
with the interplay between fluid flow, electric stresses and interfacial tension leading to droplet structures and dynamics~\cite{Vlahovska2019}. As the electric field driving frequency is lowered, the system transitions from classical ER-like chain formation at high frequencies to more complex, chaotic or turbulence-like states at low frequencies and dc fields~\cite{varshney2012self,varshney2016multiscale,Tadavani2017,bahraminasr2026electrorheoimaging,bahraminasr2026tunable}. 

It has recently been found that mesoscale heterogeneity emerges under combined shear and electric field, with banded structures consisting of alternating droplet-rich and droplet-depleted regions observed under steady shear \cite{bahraminasr2026electrorheoimaging}. 
Rheology is indeed a powerful tool for characterizing such complex fluids~\cite{cao2006structure}; however, 
conventional bulk stead-shear rheology averages over the entire sample volume and may not fully capture the microscopic dynamics in systems that develop spatial heterogeneity. Small-amplitude oscillatory shear can provide a gentler perturbation; 
however, microrheological approaches such as differential dynamic microscopy (DDM)~\cite{cerbino2008differential,edera2017differential} provide a complementary perspective to macroscopic rheology by probing local viscoelasticity non-invasively through tracer dynamics.

A rheological feature of such non-equilibrium steady states is hysteresis, commonly attributed to the competition between structure formation and shear-induced breakdown, indicating that ER fluids retain a history (or ``memory'') of prior driving through their evolving microstructure \cite{kim2006hysteresis,han2003hysteresis,aizawa2001hysteresis,keim2019memory}. 
Unlike equilibrium systems, which lose memory of prior states upon relaxation, driven and disordered materials can retain signatures of past perturbations that can be recovered through appropriate protocols.
Several forms of memory have been identified across soft matter systems. These include memory of driving direction, observed in sheared suspensions and granular materials \cite{gadala1980shear,toiya2004transient}, and memory of maximum applied input, as in the Kaiser and Mullins effects \cite{kaiser1950untersuchungen,mullins1948effect}. More complex behavior arises in systems that encode temporal history, such as the Kovacs effect, where evolution depends on prior aging \cite{kovacs2006transition,mossa2004crossover}. Memory is thus a unifying feature of driven materials, yet its manifestation depends strongly on the underlying microstructure and dynamics.
In relation to the electrically driven emulsion, to what extent do macroscopic rheological measurements reflect the underlying microscopic dynamics in a system with strong spatial heterogeneity? Furthermore, when the driving (e.g., electric field) is removed, how rapidly does the system lose memory of its driven state, and on what timescales do repeated field cycles become independent?

In this work, we employ DDM microrheology to connect microscopic dynamics to macroscopic viscoelastic moduli via the generalized Stokes–Einstein relation. %(Sec.~\ref{sec:ddm_theory}). 
To address questions of structure–property relations in electrically driven emulsions, we combine it with oscillatory rheology %\cite{bahraminasr2026electrorheoimaging}
to correlate bulk mechanical response with microscopic relaxation dynamics. We also probe memory effects using time-resolved electrorheoimaging (ERI) to examine the evolution of spatial structures under intermittent electric fields.

\section{Experimental Methods}
\subsection{Materials}
\label{sec:material}
In this experiment, we used an emulsion of castor oil in motor oil (which we denote [Ca:Mo]), specifically NOW Solutions castor oil and Irving MAX1 10W-30 motor oil with a volume ratio of 1:10. The electrical conductivity %($\sigma_E$) 
of castor oil (motor oil) was measured at $3.96\pm0.01$ picomho/cm ($102\pm1$ picomho/cm), the relative permittivity ($\kappa$) at $5.9\pm1.3$ ($2.58\pm0.52$), and the viscosity ($\eta$) at  $0.961\pm0.002$ Pa.s ($0.149\pm0.002$ Pa.s), respectively. 

\subsection{Experimental setup}

\begin{figure}[htp!]
    \centering
    \includegraphics[width=1\linewidth]{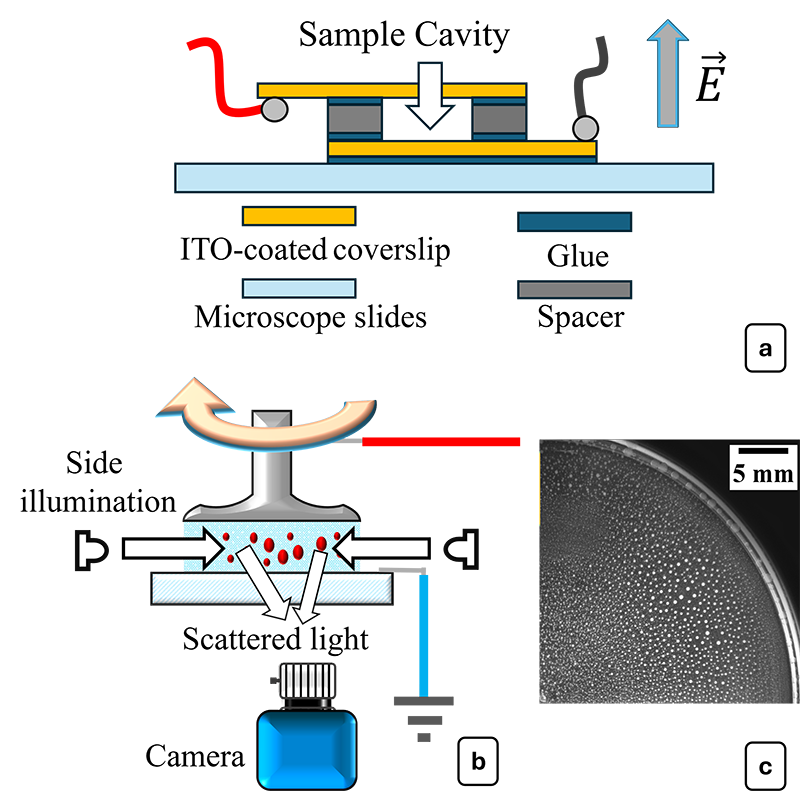}
    \caption{{\bf Experimental Setup} (a) Schematic of the vertical field sample cell consisting of two parallel indium tin oxide (ITO)–coated glass coverslips separated by a $270\mu$m spacer. (b) Schematic of electro-rheo-imaging (ERI) setup. Electrical contact is made with the high-voltage wire (red and blue) the sample illuminated from the side. Imaging is performed from the bottom with a camera. (c) A sample image of bright droplets against dark background.}
    \label{fig:setups}
\end{figure}
Sample cells (Fig.~\ref{fig:setups}) consisted of two parallel indium tin oxide (ITO)–coated glass coverslips separated by a 270$\mu$ m spacer. The chamber was filled with the [Ca:Mo] emulsion and then sealed. The chosen thickness serves two purposes: first, it is large enough that it does not suppress out-of-plane droplet motion \cite{tadavani2016effect}; second, our assembly method reproducibly yields a consistent cell thickness of 270 ± 5 $\mu$m. In this configuration, the applied electric field is parallel to the optical axis, creating what is referred to as a vertical field cell. The chosen thickness serves two purposes: first, it is large enough that it does not suppress out-of-plane droplet motion \cite{tadavani2016effect}; second, our assembly method reproducibly yields a consistent cell thickness of 270 ± 5$\mu$m.
A low–voltage input ($<12$~V) was provided by either a DC supply (Extech 382270) or an AC function generator (Tektronix AFG~3022), then amplified $200\times$ by a Trek~2220 high–voltage amplifier. Inputs exceeding 12~V saturate the amplifier to a constant output. The output was monitored via the amplifier’s monitor port and measured by a voltmeter (DC) or oscilloscope (AC). %The chamber was filled with a [Ca:Mo] emulsion. 

Figure~\ref{fig:setups}(b) shows the ERI setup: details were provided in previous work~\cite{bahraminasr2026electrorheoimaging}. The system is built around a stress-controlled rotational rheometer (Anton Paar MCR 301). A 50 mm stainless-steel parallel-plate probe serves as the top electrode, while the bottom plate is replaced with an ITO-coated glass disk. Imaging was performed using a PCO Edge 4.1 camera equipped with a macro lens to capture diffracted light from the sample. The dispersed phase was dyed to enhance contrast, producing bright droplets against a dark background (Fig.~\ref{fig:setups}(c)).

\subsection{Macroscopic rheology}

Bulk rheological measurements were performed with the MCR 301 rheometer, with a parallel-plate geometry of radius $R = 25\,\mathrm{mm}$. The sample was loaded into the ERI cell, which uses an ITO-coated disk as lower plate to allow the simultaneous application of an electric field perpendicular DC  to the shear plane.
Small-amplitude oscillatory shear experiments were conducted in the linear viscoelastic regime by imposing a sinusoidal strain
\begin{equation}
\gamma(t) = \gamma_{0}\sin(\omega t),
\end{equation}
with strain amplitude $\gamma_{0}=5\%$.
The resulting stress response was used to determine the complex shear modulus
\begin{equation}
G^{*}(\omega)=G'(\omega)+iG''(\omega),
\end{equation}
where $G'(\omega)$ is the storage modulus and $G''(\omega)$ is the loss modulus.
Frequency sweeps were performed over the range $\omega = \mathrm{1}$--$\mathrm{200}\,\mathrm{rad/s}$ at a controlled temperature of $T = 20^\circ\mathrm{C}$.

\subsection{\label{ddm_method}DDM-based microrheology}

\begin{figure*}[th!]
\centering
\includegraphics[width=0.95\linewidth]{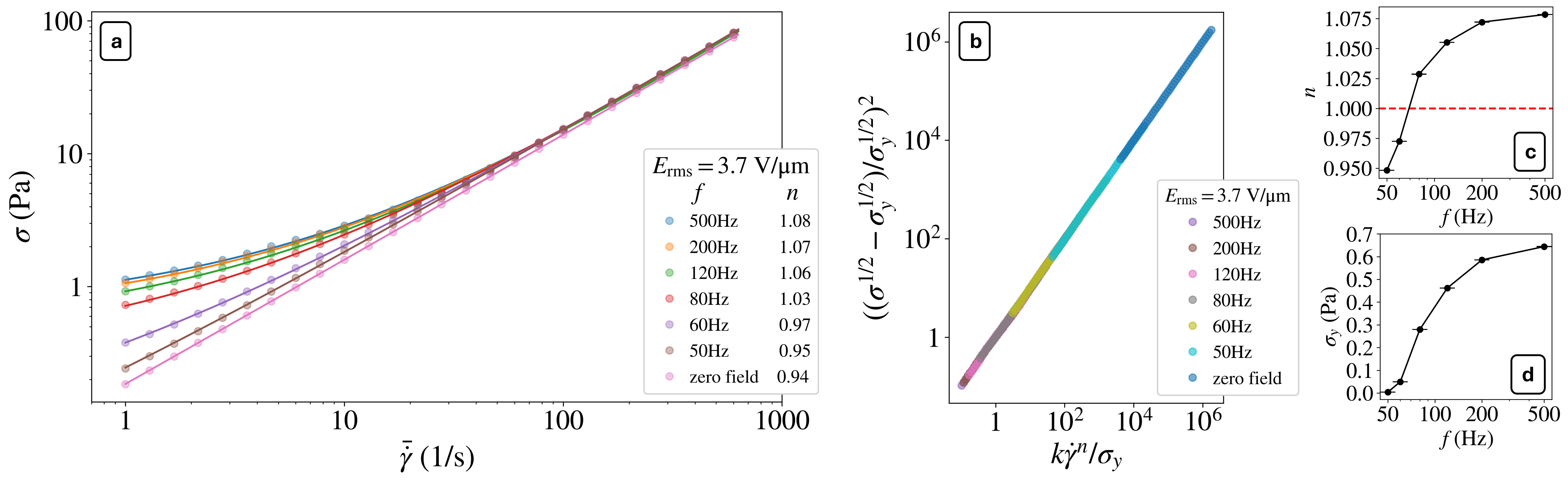}
\caption{
{\bf Frequency-dependent flow curves: curve-fitting and data collapse.}
(a) Flow curves measured at fixed root-mean-square electric field, $E_{\mathrm{rms}} = 3.7~\mathrm{V/\mu m}$, for frequencies between 50 and 500~Hz, fitted using gHB model in Eq.~\ref{eq:HBnC2}. The fitted shear-thinning exponent remains close to $n=1$
(b) Master-curve representation of the data in (a), obtained by plotting the reduced shear stress $(\sigma^{1/2} - \sigma_y^{1/2})/\sigma_y^{1/2})^2$ as a function of the rescaled shear rate $\frac{k \dot{\gamma}^{n}}{\sigma_y}.$ A good collapse is observed, with small deviations at the lowest frequencies. (c) and (d) Evolution of the fitted parameters $n$ and $\sigma_y$ as a function of frequency. Both parameters increase from their low-frequency values (50 Hz) and gradually approach a plateau at higher frequencies, consistent with a saturation of the system response. Error bars are smaller than marker size.}
\label{fig:ER_fs}
\end{figure*}
Microscopic viscoelastic properties were measured using differential dynamic microscopy (DDM), which provides access to tracer particle dynamics without the need for explicit trajectory tracking \cite{cerbino2008differential,edera2017differential}.
Time-resolved image sequences were acquired under the same field and shear conditions as the bulk rheology experiments.
DDM quantifies dynamics through the image structure function
\begin{equation}
D(\mathbf{q},t)
= \left\langle \left| I(\mathbf{q};t_{0}+t)-I(\mathbf{q};t_{0}) \right|^{2} \right\rangle_{t_{0}},
\label{eq:Dqt}
\end{equation}
which can be expressed in the standard form
\begin{equation}
D(\mathbf{q},t)
= A(\mathbf{q})\left[1-f(\mathbf{q},t)\right]+B(\mathbf{q}),
\label{eq:Dqt_fit}
\end{equation}
where $A(\mathbf{q})$ is the amplitude, $B(\mathbf{q})$ accounts for experimental noise, and $f(\mathbf{q},t)$ is the intermediate scattering function.

For tracer particles undergoing thermally driven motion with approximately Gaussian displacement statistics, the mean-squared displacement (MSD) can be obtained directly from the measured structure function as \cite{edera2017differential}

\begin{equation}
\langle \Delta r^{2}(t)\rangle
= -\frac{4}{q^{2}}
\ln\!\left(
1-\frac{D(\mathbf{q},t)-B(\mathbf{q})}{A(\mathbf{q})}
\right).
\label{eq:MSD_from_Dqt}
\end{equation}
To determine $A(q)$ and $B(q)$ without assuming a specific functional form for $f(q,t)$, we employed the model-free DDM procedure~\cite{edera2017differential}. 
This approach is based on a self-consistency requirement: the true MSD must be independent of the chosen wavevector $q$, such that MSD curves computed from different $q$ values collapse onto a single master curve. 
The MSD is then related to the viscoelastic response $G^{*}(\omega)$ through the generalized Stokes--Einstein relation (GSER),
\begin{equation}
G^{*}(\omega) = \left. \frac{d\,k_B T}{3\pi a\, s\, \widetilde{\langle \Delta r^2\rangle}(s)} \right|_{s = i\omega},
\end{equation}
where $d$ is the dimensionality, $a$ is the tracer radius, and $\widetilde{\langle \Delta r^2\rangle}(s)$ denotes the Laplace transform of the MSD \cite{edera2017differential}.

\section{Emulsion at high field frequencies behaves as an electrorheological fluid}
Measurements are carried out at a fixed root-mean-square electric field $E_{\mathrm{rms}} = 3.7~\mathrm{V/\mu m}$, while the frequency is varied. Previous work \cite{bahraminasr2026electrorheoimaging} has shown that increasing frequency enhances dipolar interactions.  
In Fig.~\ref{fig:ER_fs}(a), flow curves, plotted for $f=50$ to 500 Hz, converge at high shear rate ($\dot{\gamma}$) but deviate from each other at lower $\dot{\gamma}$. 

In classical electrorheological (ER) behavior~\cite{winslow1949induced,parthasarathy1996electrorheology}, the effect of an electric field is to give rise to a yield stress $\sigma_{y}$ that increases with field intensity; such a deviation would result in the differing shear stresses at low $\dot{\gamma}$. For a non-Newtonian fluid, one can describe this
using what we refer to as a generalized Herschel--Bulkley (gHB) model,
%\begin{equation}
$\sqrt{\sigma(\dot{\gamma})} =  \sqrt{\sigma_{y}} + \sqrt{k \dot{\gamma}^{n}} $,
%\label{eq:HBnC}
%\end{equation}
or equivalently,
\begin{equation}
\sigma(\dot{\gamma}) = \sigma_{y} + k \dot{\gamma}^{n} + 2\sqrt{k \sigma_{y}\dot{\gamma}^{n}}.
\label{eq:HBnC2}
\end{equation}
This form is motivated by the Casson model~\cite{macosko1994rheology}, with the third term capturing a smooth transition between a low-$\dot{\gamma}$ yield-stress and Newtonian behavior at higher $\dot{\gamma}$.
Here, $k$ is an effective consistency parameter and $n <1 $ ($n = 1$) for a shear-thinning (Newtonian) fluid. 

\begin{figure*}[htb]
\centering
\includegraphics[width=0.95\linewidth]{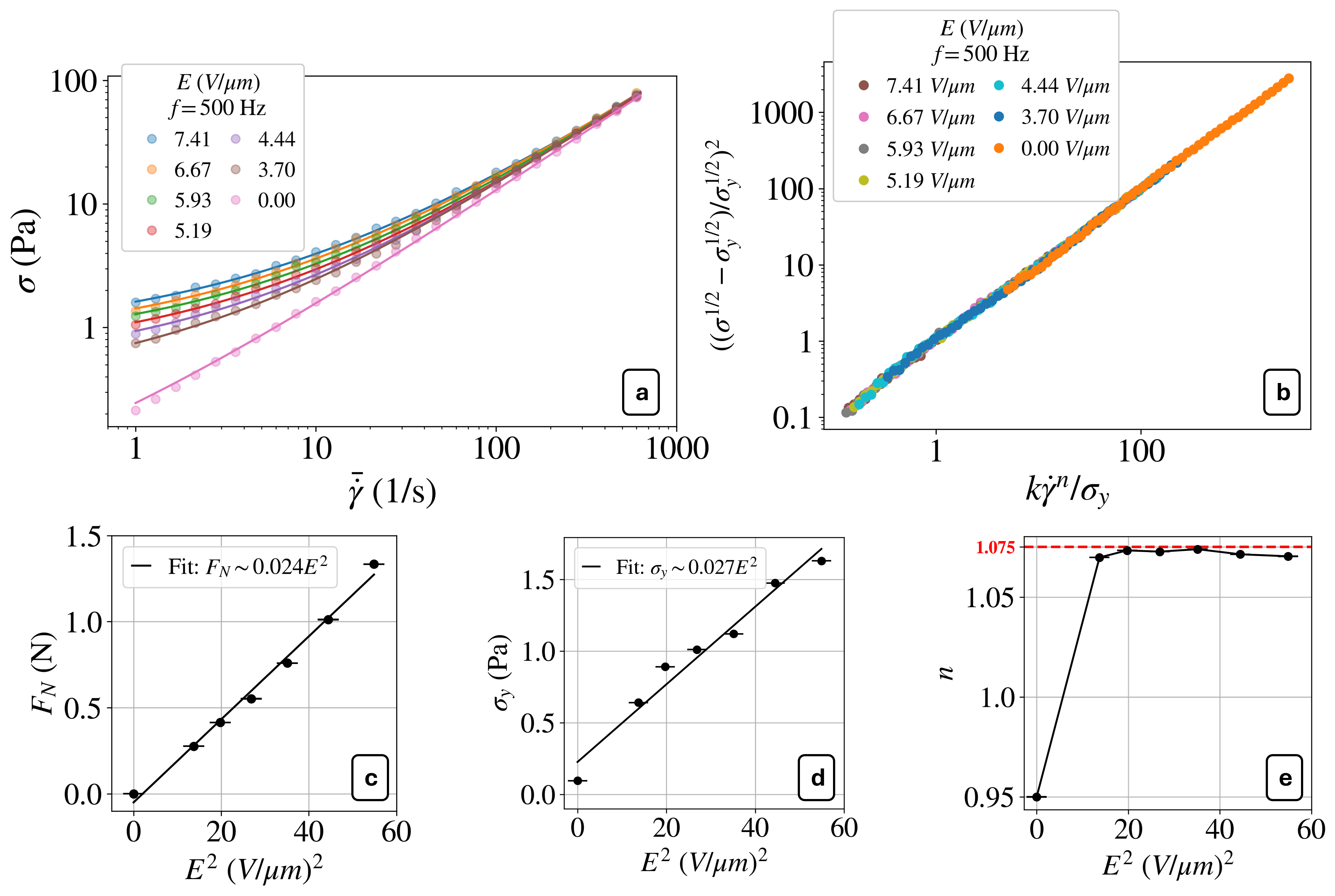}
\caption{
{\bf Field-amplitude-dependent flow curves: curve-fitting and data collapse.}
(a) Flow curves measured at fixed frequency, $500~\mathrm{Hz}$, for different electric field strengths, showing that the Casson limit ($n=1$) provides a good description over the full range of applied fields. (b) The corresponding rescaled data collapse onto a single master curve. 
(c) Measured normal force as a function of $E_{\mathrm{rms}}^{2}$. The approximately linear dependence is consistent with the capacitor scaling $F_N = \bar{\kappa}\epsilon_0 A E^2/2$, and the fitted effective dielectric constant is close to the expected value for the emulsion.
(d) Fitted yield stress $\sigma_y$ as a function of $E_{\mathrm{rms}}^{2}$. The approximately linear increase indicates that the field-induced stress scale is set by Maxwell stresses and grows with increasing electric field strength.
}
\label{fig:ER_Es}
\end{figure*}
\subsection{Collapse of frequency-dependent flow curves}

The results in Figure~\ref{fig:ER_fs}(a) are fit using the gHB model in Eq.~\ref{eq:HBnC2}. 
The fitted exponent $n$ is found to be close to unity, but $n$ is not 1, crossing from mildly shear thinning at 50Hz to shear thickening for 80 and above. We can collapse all data by defining a reduced shear stress
\begin{equation}
\Delta \sigma 
= \left( \frac{\sigma^{1/2} - \sigma_y^{1/2}}{\sigma_y^{1/2}} \right)^2,
\end{equation}
and a rescaled shear rate 
\begin{equation}
\frac{k \dot{\gamma}^{n}}{\sigma_y}.
\end{equation}
With this rescaling, all data collapse onto a single master curve (Fig.~\ref{fig:ER_fs}(b)) that is linear, with slope of unity: this simply highlights the goodness of fit to the gHB model.
The power law $n$ varies from 0.95 to 1.08 with increasing $f$ (Fig.~\ref{fig:ER_fs}(c)), while the yield stress $\sigma_y$ increases from 0 to 0.7 Pa (Fig.~\ref{fig:ER_fs}(d)).

\subsection{Collapse of flow curves at different field strengths}

To isolate the effect of electric field strength, measurements are performed at a fixed frequency of $500~\mathrm{Hz}$ for electric field ranging from 2.96 to 7.41~$\mathrm{V/\mu m}$: these results, shown in Fig.~\ref{fig:ER_Es}(a), also show qualitatively an increase in yield stress with increasing field strength (increasing deviation at low $\dot{\gamma}$). Equation \ref{eq:HBnC2} again provides a good fit across all field strengths, and once again the data collapse onto a single master curve using the same rescaling (Fig~\ref{fig:ER_Es}(b)).
The measured normal force (Fig. ~\ref{fig:ER_Es}(c)) is found to increase linearly with the electric field intensity $E_{\mathrm{rms}}^{2}$. This is expected: for a parallel-plate capacitor of area $A$, the normal force is given by
\begin{equation}
    F_N = \frac{\bar{\kappa} \epsilon_0 A}{2} E^2,
    \label{eq:FN}
\end{equation}
where $\bar{\kappa}$ is the effective dielectric constant of the emulsion \cite{bahraminasr2026electrorheoimaging}. The fitted value $\bar{\kappa} = 2.85 \pm 0.11$ is consistent with the emulsion parameters (section \ref{sec:material}): $\bar{\kappa}_{\mathrm{calc}} \simeq (0.1)(5.9) + (0.9)2.58) \simeq 2.9$. 
The fitted yield stress $\sigma_y$ is examined. The yield stress (Fig.~\ref{fig:ER_Es}(d)) also increases linearly with $E^{2}$. This trend is consistent with the scaling of Maxwell stresses \cite{torza1971electrohydrodynamic}, indicating that stronger electric fields enhance interfacial stresses and droplet interactions. Together, the increase in shear stress and the $E^{2}$ scaling of both normal force and yield stress demonstrate that the electric field controls both the tangential and normal components of the emulsion stress.

\section{The low-frequency EHD regime is dominated by convection}
\begin{figure}[htp!]
\centering
\includegraphics[width=0.81\linewidth]{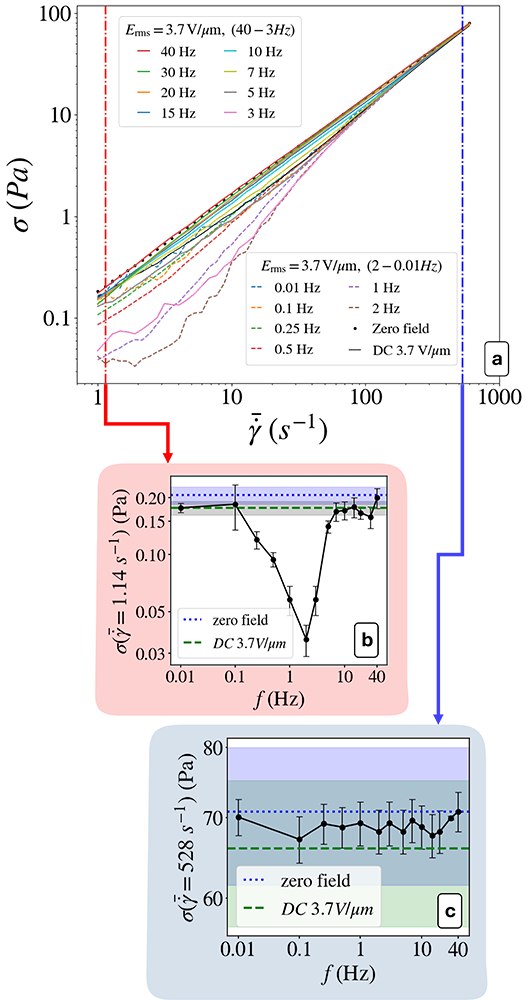}
\caption{\textbf{Failure of collapse in the EHD regime:}
(a) Flow curves showing non-monotonic evolution of the shear stress $\sigma$ with decreasing frequency in the EHD-dominated regime. Curves are shifted downward from 40 Hz to $\sim 2$ Hz, and then upwards from 2 Hz to DC.
(b,c) Shear stress at fixed shear rates, (b) $\bar{\dot{\gamma}} = 1.14~\mathrm{s^{-1}}$ and (c) $528~\mathrm{s^{-1}}$, highlight the contrasting behavior: strong frequency dependence at low shear (EHD-dominated) and convergence of all curves at high shear where hydrodynamic forcing dominates.}
\label{<2Hz_vis}
\end{figure}

Measurements were also carried out in the DC- and low-frequency regime (DC to 40 Hz in Fig.~\ref{<2Hz_vis}(a)). In contrast to the ER regime, the low-frequency response is not monotonic: the curves shift downward between 40 Hz and 1 Hz, and then back upward from 1Hz to DC. 
As a result, attempts to collapse the flow curves in the low-frequency (EHD-dominated) regime are unsuccessful  Applying the same Casson-type rescaling does not yield a master curve: the datasets remain systematically separated, and the deviations cannot be accounted for by simple adjustments of $k$ or $\sigma_y$.

As shown in previous work~\cite{bahraminasr2026electrorheoimaging}, decreasing the frequency alters the microstructure; below $\sim 10$ Hz, droplet breakup becomes dominant. In addition, convection rolls with a characteristic period slightly above $1~\mathrm{s}$ lead to a consistent reduction in shear stress near this frequency, effectively assisting the imposed shear~\cite{bahraminasr2026electrorheoimaging}.

The shear stress $\sigma$ at $\bar{\dot{\gamma}}=1.14~\mathrm{s}^{-1}$ and at $\bar{\dot{\gamma}}=528~\mathrm{s}^{-1}$ are shown in Fig.~\ref{<2Hz_vis}(b) and (c), respectively. In Fig.~\ref{<2Hz_vis}(b), where EHD forces dominate, the shear stress decreases as the frequency is reduced toward $\sim 2$ Hz, and increases again at lower frequencies. In contrast, in Fig.~\ref{<2Hz_vis}(c), where shear dominates, the curves converge across all frequencies, as expected.

This lack of collapse indicates that multiple coupled mechanisms determine the shear stress response (Fig.~\ref{<2Hz_vis}). Unlike the ER case, where dipolar interactions define a well-characterized yield stress, the EHD regime exhibits continuously evolving, flow-coupled structures. Remarkably, this results in a non-monotonic dependence at low shear rates on the field frequency $f$, with a minimum that is very close to the emergent convection roll frequency $\nu$.

\section{Recurrence and divergence in EHD banding}
%===============================================================
%===============================================================

\label{sec:eri_analysis}

\begin{figure*}[hpt!]
\centering
\includegraphics[width=0.8\textwidth]{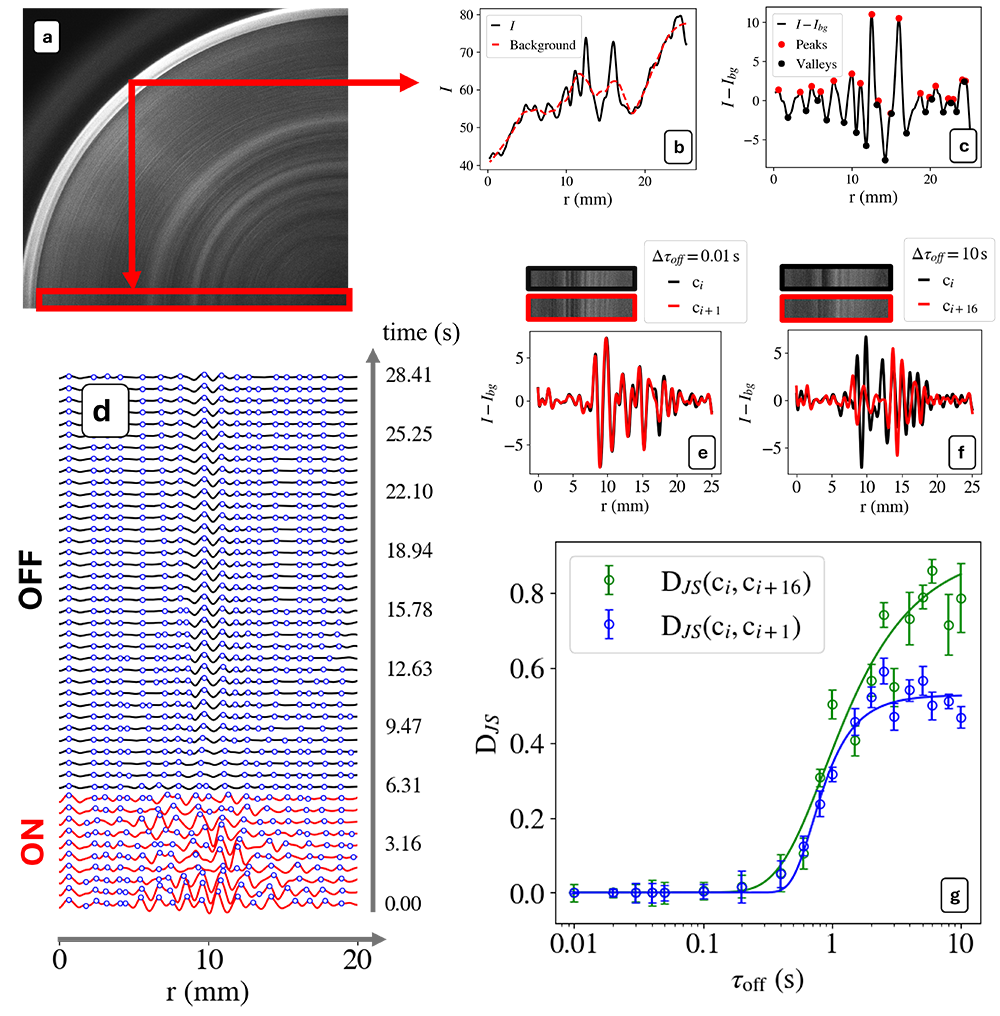}
\caption{{\bf Band identification via radial intensity analysis:} (a) The full field of view of aand the extracted $1000\times50$ pixel strip used for analysis.
(b) One-dimensional radial intensity profile $I(r,t)$ obtained by averaging a 10-pixel-wide band, together with the smoothed background profile $I_{\mathrm{bg}}$ (200-pixel moving average).
(c) Fluctuation signal $\Delta I(r,t)=I(r,t)-I_{\mathrm{bg}}$, where local maxima identify droplet-rich bands and local minima correspond to depletion regions. {\bf Recurrence as a function of the $\tau_{\mathrm{off}}$ timescale:}
Analysis of band formation and relaxation under intermittent electric field.
(d) Single electric field on (red curves) and off (black curves) cycle: Temporal decay of band contrast immediately after switching off the electric field (transition from red to black curves), showing a rapid initial decrease followed by a steady state. The shear rate, $\bar{\dot{\gamma}} = 600~\mathrm{s}^{-1}$, is maintained throughout the experiment, both during field-on and field-off conditions.
(e,f) Normalized radial intensity fluctuations $\Delta I(r)$ after 10~s of forcing at $3.7~\mathrm{V/\mu m}$ for pause times $\tau_{\mathrm{off}}=0.01~\mathrm{s}$ (e) and $\tau_{\mathrm{off}}=10~\mathrm{s}$ (f).
Black and red curves correspond to two consecutive on and off cycles ($\mathrm{c}_i$ and $\mathrm{c}_{i+1}$), with representative microscopy snapshots shown with matching borders. 
For short pauses the band profiles remain strongly correlated, whereas for longer pauses the similarity decreases.
(g) Jensen--Shannon divergence ($\mathrm{D}_{JS}$) between intensity profiles as a function of pause time, quantifying structural decorrelation. Small values indicate strong similarity between cycles, while larger values reflect progressive loss of memory of the driven state.}

\label{fig:BandsINFO}
\end{figure*}

%===============================================================
%===============================================================
Using electrorheoimaging (ERI), we see that electrically driven emulsions develop banded structures when a low-frequency or DC electric field is applied under steady shear. This was reported in detail in previous work~\cite{bahraminasr2026electrorheoimaging}. 
These bands consist of alternating droplet-rich and droplet-depleted regions (shown as bright and dark circular bands in Fig.~\ref{fig:BandsINFO}(a)) and are not observed when shear or electric field is applied independently.  
Their formation, therefore, reflects the coupled action of electrohydrodynamic forcing and shear-induced organization \cite{bahraminasr2026electrorheoimaging}.

Here, we examine how likely bands locations will recur when the field is turned off and on again: this recurrence can give some insight into memory in the system. To quantify the spatial evolution of the banded state, we analyze the radial intensity distribution. 
Each frame is a $1000\times50$ pixel strip. The entire image and the strip that is used for analysis is shown in Fig.~\ref{fig:BandsINFO}(a).
Within this strip, a 10-pixel-wide band extending from the center toward the edge is averaged to obtain a one-dimensional intensity profile $I(r,t)$, such as is shown in Fig.~\ref{fig:BandsINFO}(b). 
The profile is then smoothed using a 200-pixel moving average to obtain $I_{bg}$ (dashed red line in Fig.~\ref{fig:BandsINFO}(b)), and the fluctuation signal, 
$I(r,t)-I_{bg}$, shown in Fig.~\ref{fig:BandsINFO}(c), displays
local maxima corresponding to droplet-rich bands, and local minima identifying depletion regions.

During imaging, a constant average shear rate of $\bar{\dot{\gamma}}=600~\mathrm{s^{-1}}$ is applied while the electric field is switched on and off. 
Upon initial application of the field, the band pattern develops gradually from an initially uniform droplet distribution and fully developed within approximately 5~s. 
The electric field is generated using a National Instruments PCI-4462 card controlled via NI-DAQmx API for Python, producing a square-wave signal with a fixed on-time of 10~s and variable off-time. 
Right before the end of each field-on interval, the camera is triggered at 500~fps to capture one full revolution.

Once a steady banded state is established, switching off the field leads to a gradual reduction in spatial contrast as the droplet distribution relaxes toward a homogeneous (unbanded) state. Figure~\ref{fig:BandsINFO}(d) shows the final $\sim$6~s of the field-on state (red curves), during which the bands are fully developed but still evolving. Immediately after the field is switched off (black curves), the band contrast decays rapidly before approaching a steady residual value. While some peaks remain, they are stationary and reflect persistent density inhomogeneities rather than dynamically evolving bands.

Figure~\ref{fig:BandsINFO}(e) shows normalized intensity fluctuations as a function of radial position, obtained after 10~s of forcing at $3.7~\mathrm{V/\mu m}$, for two consecutive cycles ($\mathrm{c}_i$ in black and $\mathrm{c}_{i+1}$ in red), with an ``off'' time of $\tau_{\mathrm{off}}=0.01~\mathrm{s}$. Also shown are the snapshots from which these $I(r,t)-I_{bg}$ profiles are obtained (with black and red border respectively).  Figure~\ref{fig:BandsINFO}(f) shows the same for a much longer off time $\tau_{\mathrm{off}}=10~\mathrm{s}$. 
It is apparent that for short off times, the profiles remain strongly correlated, whereas for longer pauses the correlation decreases. 
The relaxation dynamics are quantified using the Jensen–Shannon (JS) divergence, $\mathrm{D}_{JS}$, between intensity profiles at different times. This specific metric is chosen because it provides a symmetric, always well-defined, and strictly bounded measure of similarity between spatial distributions, overcoming the asymmetry and potential unboundedness of the standard Kullback–Leibler divergence \cite{nielsen2019jensen}. This bounded metric provides a measure of structural decorrelation: small values indicate similar band patterns, while larger values reflect progressive loss of memory of the driven state (Fig.~\ref{fig:BandsINFO} (g)).

Finally, we examine the reproducibility of band formation under intermittent forcing using a square-wave protocol. 
For each delay interval between $0.01~\mathrm{s}$ and $10~\mathrm{s}$, the experiment is repeated 19 times. 
$\mathrm{D}_{JS}$ is calculated between consecutive cycles (blue curve) and between cycles separated by 16 intervals (green curve). $\mathrm{D}_{JS}$ is zero when two signals are identical, and approaches 1 for very dissimilar signals. 
The field is applied for 10~s, while the off-time varies between $0.05$ and $10~\mathrm{s}$. 
The results (Fig.~\ref{fig:BandsINFO}) show that the system rapidly loses memory after the field is removed. 
A significant increase in $\mathrm{D}_{JS}$ occurs within approximately 0.5~s, and the signal approaches a plateau after 1–2~s. This timescale is remarkably consistent with the turnover time of the electroconvective rolls. Independent evidence for such rolls is provided by microscopy in the vertical-field cell, where droplets move in and out of focus over 70 ms, demonstrating out-of-plane circulation~\cite{bahraminasr2026electrorheoimaging}. In addition, the spacing of the diffuse EHD bands at onset scales with the gap as $d/h = 2.06 \pm 0.15 \approx 2$, consistent with adjacent counter-rotating rolls whose lateral periodicity is set by the cell thickness. Taken together, these observations support the interpretation that the characteristic timescale extracted here reflects electroconvection rather than an unrelated structural relaxation process. The similarity between consecutive-cycle and 16-cycle-separated measurements indicates that, for off times longer than about 1~s, banding events become statistically independent.

\section{\label{sec:moduli}Examining emulsion elasticity requires a gentle probe}
An important question in ERI is whether the microscopic (droplet-scale) imaging and the macroscopic rheological signatures are reporting on the same phenomena. As an extreme case, edge artifacts in rheology could affect the latter signature profoundly while not showing up in images of the interior of the sample cell. We test this by looking for consistency between macroscopic rheology and microrheology. These measurements were carried out for a 1:10 castor oil–in–motor oil emulsion under an applied DC electric field of 3.7~$V/\mu m$.

To quantify the viscoelastic response of electrically driven emulsions across both macroscopic and microscopic length scales, the storage modulus $G'(\omega)$ and loss modulus $G''(\omega)$ have been measured using two complementary approaches. First, we carry out conventional oscillatory shear rheology in DC, which probes the bulk mechanical response. Then we employ DDM-based microrheology, which infers viscoelastic moduli from the microscopic images.

\begin{figure}[hpt!]
\centering
\includegraphics[width=0.4\textwidth]{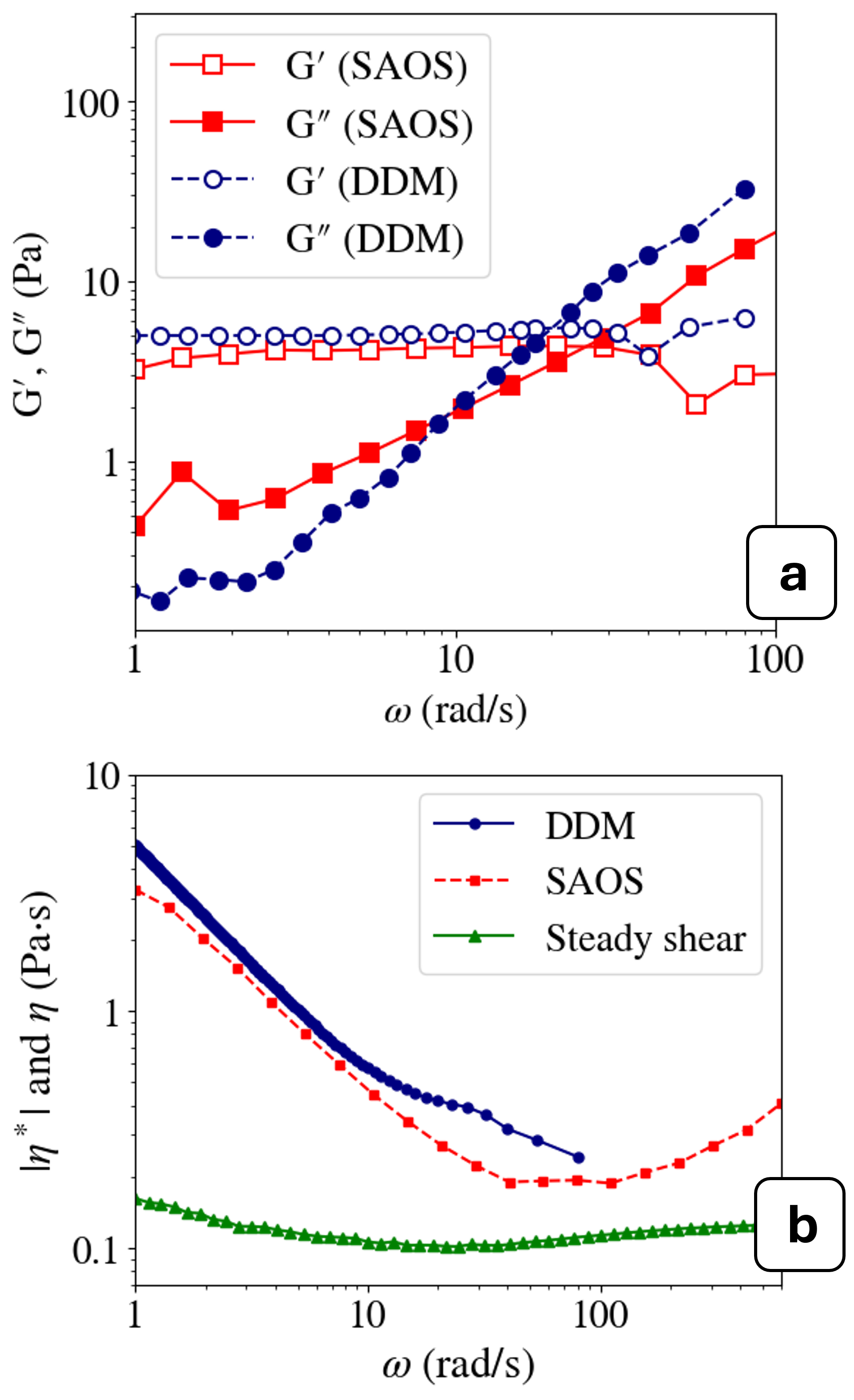}
\caption{
%Comparison of 
{\bf Comparing macro- and micro-rheology.}
(a) Storage ($G'$) and loss ($G''$) moduli obtained from macroscopic oscillatory rheology (dashed lines) and DDM microrheology (solid lines) at 3.7~$V/\mu m$ (DC). In both measurements, $G'(\omega)$ exceeds $G''(\omega)$ over most of the accessible angular frequency range, indicating predominantly elastic behavior under the applied field. The overall agreement between the two techniques demonstrates that both probe the same field-induced viscoelastic state, while differences at higher frequencies reflect the distinct nature of fluctuation-based and deformation-based measurements. 
(b) Complex viscosity $|\eta^{*}|$ as a function of angular frequency $\omega$, obtained from SAOS and from DDM-based microrheology, is compared with steady-shear measurements. The empirical Cox--Merz rule is seen not to be valid. The SAOS and DDM results show comparable magnitudes, whereas the viscosity inferred from the flow curve is substantially lower, indicating that steady shear disrupts the field-induced microstructure more strongly than SAOS or DDM.}
\label{fig:GpGpp}
\end{figure}

The viscoelastic moduli obtained from bulk small-amplitude oscillatory shear (SAOS, blue symbols) and  microrheology (red symbols) are compared directly in Fig.~\ref{fig:GpGpp}(a). During the oscillatory experiments, no band formation is observed, and both microscopic and macroscopic measurements are performed under uniform sample conditions. The overall behavior obtained from the two methods is qualitatively similar. 
In both datasets, $G'(\omega)$ exceeds $G''(\omega)$ over most of the accessible frequency range, indicating that the field-driven emulsion exhibits predominantly elastic behavior on these timescales. The weak frequency dependence of $G'(\omega)$ suggests a broad distribution of relaxation times associated with electrically induced droplet polarization and interfacial interactions, which impart solid-like characteristics without complete dynamical arrest.

The two techniques show comparable trends in $G'(\omega)$, supporting the conclusion that both probe the same underlying field-induced viscoelastic state. Quantitative differences are more pronounced in $G''(\omega)$, particularly at higher frequencies where the rheological $G''$ increases more steeply and approaches (or exceeds) $G'$. This enhanced dissipation in bulk rheology is consistent with the fact that oscillatory measurements impose mechanical deformation, potentially introducing additional dissipative mechanisms such as interfacial rearrangements, shear-induced microstructural reorganization, or boundary effects. In contrast, DDM infers moduli from spontaneous tracer fluctuations and therefore emphasizes intrinsic microscopic relaxation dynamics under the applied field with minimal mechanical perturbation.

%\mb{\subsection{Three methods for measuring viscosity}

Flow curve measurements with a rheometer provide direct access to the shear-rate-dependent response of the material, but the applied steady shear has the highest potential to perturb the sample microstructure. This effect is evident in Fig.~\ref{fig:GpGpp}(b), where the absolute value of the complex viscosity $|\eta^{*}|$ (SAOS in blue, DDM in red) is plotted as a function of angular frequency $\omega$ and compared with the steady-shear flow curve (in black). If a Cox-Merz-type relationship were valid, these values would be roughly similar and show similar trends.
The apparent viscosities obtained from SAOS and DDM are in relatively good agreement, whereas the viscosity inferred from the flow-curve measurement is substantially lower. This difference highlights the importance of microstructural disturbance during measurement. DDM-based microrheology is non-invasive, and SAOS only weakly perturbs the structure in the linear-response regime. In contrast, steady flow can strongly disrupt the field-induced structures, leading to a pronounced reduction in the measured viscosity.

\section{Discussion}

The results presented here establish a coherent picture of electrically driven emulsions across two distinct regimes, and clarify how macroscopic rheology, microscopic dynamics, and spatial structure are connected in each case.

In the high-frequency electrorheological (ER) regime, the emulsion behaves as a field-tunable yield-stress fluid. The successful description of the flow curves using a generalized Herschel--Bulkley (Casson-type) form, together with the collapse onto a master curve, indicates that the rheology is governed by a single dominant stress scale. The observed scaling $\sigma_y \sim E^2$ and the corresponding $F_N \sim E^2$ dependence are consistent with Maxwell stresses arising from dipolar interactions between polarized droplets. In this regime, the microstructure iss homogeneous and dominated by field-induced interactions rather than flow-induced rearrangements. As a result, the system admits a reduced description in terms of an effective yield stress and viscosity, similar to classical ER fluids.

In contrast, the low-frequency electrohydrodynamic (EHD) regime exhibits fundamentally different behavior. Here, the stress at low shear rate (a quantity similar to the yield stress) does not does not increase steadily with $f$, but is markedly non-monotonic, with a sharp minimum at $f \sim 2$ Hz, close the convection roll frequency $\nu \sim 1$ Hz. 
Instead, the system is characterized by continuously evolving, flow-coupled structures arising from the interplay between electric stresses, interfacial deformation, and fluid motion. The emergence of banded states under combined shear and electric field highlights the inherently heterogeneous nature of this regime. These structures cannot be captured by a simple constitutive model such as a yield-stress fluid.

There is agreement, in the EHD regime,  between macroscopic small-amplitude oscillatory shear (SAOS) and DDM-based microrheology. The response is weakly elastic ($G' > G''$) at low $\omega$. The close correspondence between $G'(\omega)$ and $G''(\omega)$ obtained from the two methods implies that both techniques gently probe the same underlying viscoelastic state. Importantly, this agreement indicates that the imposed oscillatory deformation does not significantly perturb the microstructure. 

At low field frequencies, in the presence of both steady shear and EHD forces, there is a banding instability which we refer to as EHD banding: ERI measurements provide direct insight into this spatial heterogeneity and its temporal evolution. The banded structures form rapidly under applied field, indicating that the system is highly responsive to external driving. However, once the field is removed, the structure relaxes quickly, with a characteristic memory-loss timescale of order $1~\mathrm{s}$. The use of the Jensen--Shannon divergence allows this loss of memory to be quantified in a well-defined and physically meaningful way.  
A key result is that this timescale is consistent with the turnover time of electrohydrodynamic convection rolls, suggesting that the memory of the system is not stored in static structural features, but rather in the underlying flow field that sustains the banded state. When the field is removed, the convective motion ceases, and the structure rapidly decorrelates. For sufficiently long off times, successive banding events become statistically independent, indicating complete loss of memory of prior configurations.
Taken together, these results highlight a clear distinction between the ER and EHD regimes. In the ER regime, the system behaves as a quasi-homogeneous material with well-defined rheological parameters that can be directly linked to microscopic interactions. In the EHD regime, however, the behavior is dominated by spatiotemporal dynamics, and macroscopic rheology alone is insufficient to capture the underlying physics.

This work begins to address how driven soft matter systems encode and lose memory. In electrically driven emulsions, memory is transient and dynamically maintained: it persists only while the system is actively driven, and decays on a timescale set by internal flow dynamics. This distinguishes it from other forms of memory observed in soft materials, such as those associated with irreversible structural changes or stored elastic energy. 
The combined use of rheology, microrheology, and electrorheoimaging provides a framework for linking structure, dynamics, mechanical response and memory in non-equilibrium systems. These results suggest that, in systems with strong spatial heterogeneity, a complete description requires simultaneous access to both macroscopic and local observables. Future work could explore how these ideas extend to other driven complex fluids.

\section*{Data Availability Statement}
The data that support the findings of this study are available within the article. 

\nocite{*}
\bibliography{aipsamp}
\end{document}